\documentclass[longauth]{aa}
\usepackage{graphicx}
\usepackage{txfonts}
\usepackage{xcolor}
\usepackage{hyperref}
\hypersetup{colorlinks, linkcolor=blue, citecolor=blue, urlcolor=blue}
\usepackage{amsmath}
\usepackage{ulem}
\usepackage{soul}
\usepackage{enumitem}% http://ctan.org/pkg/enumitem
\newcommand{\pivec}{\mbox{\boldmath $\pi$}}

\newcommand{\te}{t_{\rm E}}
\newcommand{\thetae}{\theta_{\rm E}}

\newcommand{\pie}{\pi_{\rm E}}
\newcommand{\pien}{\pi_{{\rm E},N}}
\newcommand{\piee}{\pi_{{\rm E},E}}
\newcommand{\dl}{D_{\rm L}}
\newcommand{\ds}{D_{\rm S}}

\def\btheta{{\vec\theta}}

\definecolor{brown}{rgb}{0.59, 0.29, 0.0}
\definecolor{darkgreen}{rgb}{0.0, 0.42, 0.24}
\definecolor{darkblue}{rgb}{0.01, 0.31, 0.59}
\definecolor{darkblue}{rgb}{0.0, 0.25, 0.42}
\definecolor{blue}{rgb}{0.0,0.0,1.0}
\definecolor{green}{rgb}{0.0,1.0,0.0}

\def\eqalign#1{\null\,\vcenter{\openup\jot
        \ialign{\strut\hfil$\displaystyle{##}$&$
        \displaystyle{{}##}$\hfil \crcr#1\crcr}}\,}
\begin{document}

%\title{Multi-planet microlensing system OGLE-2019-BLG-0468L composed of two giant planets orbiting around a G-type star}
\title{OGLE-2019-BLG-0468Lb,c: two microlensing giant planets around a G-type star}

\author{
% leading author -----------------------------
     Cheongho~Han\inst{1} 
\and Andrzej~Udalski\inst{2} 
\and Chung-Uk~Lee\inst{3} 
\and Doeon~Kim\inst{1}
\and Wei Zhu\inst{4}
\\
(Leading authors)\\
%and \\
% KMTNet ---------------------------
     Michael~D.~Albrow\inst{5}   
\and Sun-Ju~Chung\inst{3,6}      
\and Andrew~Gould\inst{7,8}      
\and Kyu-Ha~Hwang\inst{3} 
\and Youn~Kil~Jung\inst{3} 
\and Hyoun-Woo~Kim\inst{3} 
\and Yoon-Hyun~Ryu\inst{3} 
\and In-Gu~Shin\inst{3} 
\and Yossi~Shvartzvald\inst{9}   
\and Jennifer~C.~Yee\inst{10}   
\and Weicheng~Zang\inst{4}     
\and Sang-Mok~Cha\inst{3,11} 
\and Dong-Jin~Kim\inst{3} 
\and Seung-Lee~Kim\inst{3,6} 
\and Dong-Joo~Lee\inst{3} 
\and Yongseok~Lee\inst{3,11} 
\and Byeong-Gon~Park\inst{3,6} 
\and Richard~W.~Pogge\inst{8}
\and Chun-Hwey Kim\inst{12}
\and Woong-Tae Kim\inst{13}
\\
(The KMTNet Collaboration),\\
% OGLE ---------------------------
     Przemek~Mr{\'o}z\inst{2,14} 
\and Micha{\l}~K.~Szyma{\'n}ski\inst{2}
\and Jan~Skowron\inst{2}
\and Rados{\l}aw~Poleski\inst{2} 
\and Igor~Soszy{\'n}ski\inst{2}
\and Pawe{\l}~Pietrukowicz\inst{2}
\and Szymon~Koz{\l}owski\inst{2} 
\and Krzysztof~A.~Rybicki\inst{2}
\and Patryk~Iwanek\inst{2}
\and Krzysztof~Ulaczyk\inst{15}
\and Marcin~Wrona\inst{2}
\and Mariusz~Gromadzki\inst{2}          
\\
(The OGLE Collaboration)\\
% SALT team ---------------------------
     David Buckley\inst{16}              %     ---------- South African Astronomical Observatory, P.O. Box 9, Observatory 7935, Cape Town, South Africa
\and Subo Dong\inst{17}                  %     ---------- Kavli Institute for Astronomy and Astrophysics, Peking University, Beijing 100871, China     
\and Ali Luo\inst{18}      \\              %     ---------- National Astronomical Observatories, Chinese Academy of Sciences, Beijing 100012, China      
%\\
%(The SALT Observation group)\\
}

\institute{
%Institute for Astronomy (IfA), University of Vienna, T\"urkenschanzstrasse 17, A-1180 Vienna\\ \email{wuchterl@amok.ast.univie.ac.at}
%\and
%University of Alexandria, Department of Geography, ...\\ \email{c.ptolemy@hipparch.uheaven.space}
%\thanks{The university of heaven temporarily does not accept e-mails}
     Department of Physics, Chungbuk National University, Cheongju 28644, Republic of Korea  \\ \email{cheongho@astroph.chungbuk.ac.kr}                  % (1)
\and Astronomical Observatory, University of Warsaw, Al.~Ujazdowskie 4, 00-478 Warszawa, Poland                                                          % (2) 
\and Korea Astronomy and Space Science Institute, Daejon 34055, Republic of Korea                                                                        % (3)
\and Department of Astronomy and Tsinghua Centre for Astrophysics, Tsinghua University, Beijing 100084, China                                            % (4)
\and University of Canterbury, Department of Physics and Astronomy, Private Bag 4800, Christchurch 8020, New Zealand                                     % (5)
\and Korea University of Science and Technology, 217 Gajeong-ro, Yuseong-gu, Daejeon, 34113, Republic of Korea                                           % (6)
\and Max Planck Institute for Astronomy, K\"onigstuhl 17, D-69117 Heidelberg, Germany                                                                    % (7)
\and Department of Astronomy, The Ohio State University, 140 W. 18th Ave., Columbus, OH 43210, USA                                                       % (8)
\and Department of Particle Physics and Astrophysics, Weizmann Institute of Science, Rehovot 76100, Israel                                               % (9)
\and Center for Astrophysics $|$ Harvard \& Smithsonian 60 Garden St., Cambridge, MA 02138, USA                                                          % (10)
\and School of Space Research, Kyung Hee University, Yongin, Kyeonggi 17104, Republic of Korea                                                           % (11)  
\and Department of Astronomy \& Space Science, Chungbuk National University, Cheongju 28644, Republic of Korea                                           % (12)
\and Department of Physics \& Astronomy, Seoul National University, Seoul 08826, Republic of Korea                                                       % (13) 
\and Division of Physics, Mathematics, and Astronomy, California Institute of Technology, Pasadena, CA 91125, USA                                        % (14) 
\and Department of Physics, University of Warwick, Gibbet Hill Road, Coventry, CV4 7AL, UK                                                               % (15)
% -------------
\and South African Astronomical Observatory, P.O. Box 9, Observatory 7935, Cape Town, South Africa                                                       % (16)
\and Kavli Institute for Astronomy and Astrophysics, Peking University, Beijing 100871, China                                                            % (17)
\and National Astronomical Observatories, Chinese Academy of Sciences, Beijing 100012, China                                                             % (18)
%\and Astronomical Observatory, University of Warsaw, Al. Ujazdowskie 4, 00-478 Warszawa, Poland                                                          % (18)
%\\
%\email{cheongho@astroph.chungbuk.ac.kr}   
}
\date{Received ; accepted}

% \abstract{}{}{}{}{} 
% 5 {} token are mandatory
\abstract
% context heading (optional)
% {} leave it empty if necessary  
{}
% aims heading (mandatory)
{
With the aim of interpreting anomalous lensing events with no suggested models, we conducted a 
project of reinvestigating microlensing data in and before the 2019 season.  In this work, we 
report a multi-planet system OGLE-2019-BLG-0468L found from the project.
}
% methods heading (mandatory)
{
The light curve of the lensing event OGLE-2019-BLG-0468, which consists of three distinctive anomaly 
features, could not be explained by the usual binary-lens or binary-source interpretation. We find a 
solution explaining all anomaly features with a triple-lens interpretation, in which the lens is 
composed of two planets and their host, making the lens the fourth multi-planet system securely found 
by microlensing.
}
% results heading (mandatory)
{
The two planets have masses $\sim 3.4~M_{\rm J}$ and $\sim 10.2~M_{\rm J}$, and they are orbiting 
around a G-type star with a mass $\sim 0.9~M_\odot$ and a distance $\sim 4.4$~kpc.  The host of the 
planets is most likely responsible for the light of the baseline object, although the possibility 
for the host to be a companion to the baseline object cannot be ruled out.
}
% conclusions heading (optional), leave it empty if necessary 
{}


\maketitle

\section{Introduction}\label{sec:one}

Studies based on radial velocity (RV) observations have shown that about 10\% of stars have giant 
planets beyond $\sim 1$ AU \citep{Cumming2008, Fulton2021}.  How common do such cold giant planets 
have massive planetary-mass companions?  On one hand, cold giant planets have eccentric orbits that 
are generally attributed to the dynamical interactions with additional massive companions, suggesting 
that perhaps the majority of them have such companions, at least at birth \citep{Juric2008,  
Chatterjee2008}.  On the other hand, after over two decades of searches, RV surveys have only been 
able to identify the presence of massive companions to $\sim 20$ -- 30\% of known cold giants 
\citep{Wright2009, Rosenthal2021}.  This discrepancy can be potentially reduced with more discoveries 
of giant planet systems, but unfortunately RV becomes extremely inefficient in detecting planets with 
relatively lower masses and (or) longer orbital periods.

Being most sensitive to cold planets located around or beyond the water snow line, gravitational 
microlensing can play an important role in completing the demographic census of exoplanets 
\citep{Gaudi2012, Zhu2021}.  In terms of multi-planet systems, microlensing has so far detected three 
reliable two-planet systems [OGLE-2006-BLG-109L \citep{Gaudi2008, Bennett2010},  OGLE-2012-BLG-0026L 
\citep{Han2013, Beaulieu2016}, and OGLE-2018-BLG-1011L \citep{Han2019}] plus two candidate two-planet 
systems [OGLE-2014-BLG-1722L \citep{Suzuki2018} and KMT-2019-BLG-1953L \citep{Han2020a}].  Compared 
to the total number of over 100 planetary systems from microlensing, the number of multi-planet systems 
is small.  However, given the relatively low efficiency of detecting multi-planet systems with microlensing 
\citep{Zhu2014}, these numbers are already somewhat indicative that perhaps a substantial fraction of 
microlensing planets have additional planetary-mass companions \citep{Madsen2019}.

% Figure 1 ------------------------------------------------------
\begin{figure}[t]
\includegraphics[width=\columnwidth]{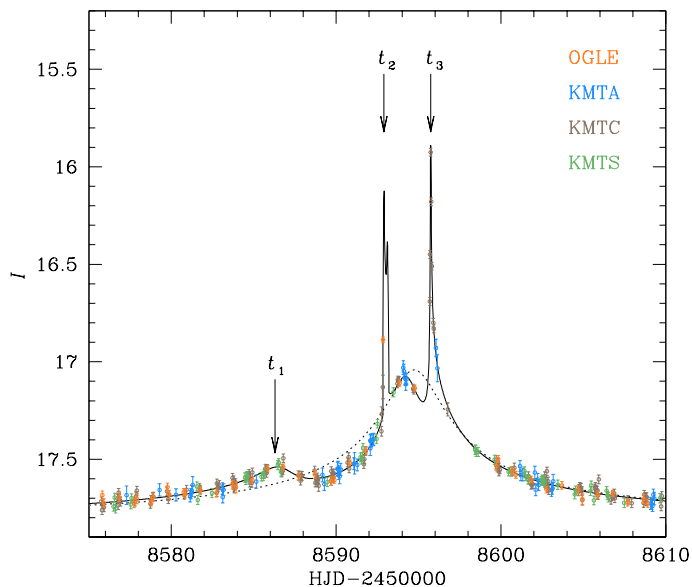}
\caption{
Light curve of the microlensing event OGLE-2019-BLG-0468.  The arrows labeled by $t_1=8586.3$, 
$t_2=8592.9$, and $t_3=8595.7$ indicate the three epochs of major anomalies. The curves drawn over the 
data points is the 1L1S (dotted curve) and 3L1S (close-close model, solid curve) models.  The zoom-in 
view of the anomaly region around $t_2$ and $t_3$ is shown in Fig.\ref{fig:two}.  
}
\label{fig:one}
\end{figure}
% --------------------------------------------------------------

Given the potential of microlensing in studying the multiplicity distribution of cold planets and 
thus the architecture of planetary systems in the cold region, it is important to detect more, secure 
multi-planet microlensing systems. This requires high-cadence observations over a large number of 
microlensing events as well as detailed light curve modelings of all anomalous events. The signal 
produced by multiple planets differs from that produced by a single planet because the individual 
planets induce their own caustics and these caustics often result in a complex pattern due to the 
interference between the caustics \citep{Danek2015a, Danek2015b, Danek2019}.  As a result, these 
signals, in most cases, cannot be described by the usual lensing models based on the binary-lens 
(2L1S) or binary-source (1L2S) interpretation. This implies that some anomalous events with signals 
produced by multiple planets are probably left unanalyzed without correct interpretations of the 
anomalies.

The amount of microlensing data was dramatically decreased during the Covid-19 pandemic, 
because many of the major survey telescopes were shut down. In order for the best use of this time, 
we have conducted a project, in which previous microlensing data collected by the KMTNet survey in 
and before the 2019 season were systematically reinvestigated. The aim of the project is to find 
events of scientific importance among those with no presented analyses. One group of events for this 
investigation are those with weak anomalies. Investigating such events led to the discoveries of 
16 microlensing planets 
including 
KMT-2018-BLG-1025Lb \citep{Han2021e}, 
KMT-2016-BLG-2364Lb, 
KMT-2016-BLG-2397Lb, 
OGLE-2017-BLG-0604Lb, 
OGLE-2017-BLG-1375Lb \citep{Han2020e}, 
KMT-2018-BLG-0748Lb \citep{Han2020d}, 
KMT-2019-BLG-1339Lb \citep{Han2020b}, 
KMT-2018-BLG-1976, 
KMT-2018-BLG-1996, 
OGLE-2019-BLG-0954 \citep{Han2021d},  
OGLE-2018-BLG-0977Lb, 
OGLE-2018-BLG-0506Lb, 
OGLE-2018-BLG-0516Lb, 
OGLE-2019-BLG-1492Lb, 
KMT-2019-BLG-0253 \citep{Hwang2021}, and
OGLE-2019-BLG-1053 \citep{Zang2021}.

% Figure 2 ------------------------------------------------------
\begin{figure}[t]
\includegraphics[width=\columnwidth]{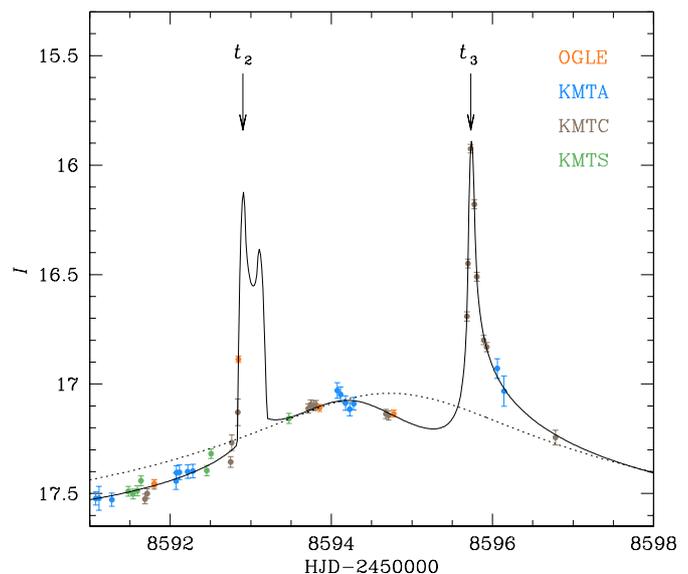}
\caption{
Zoom-in view of the anomaly region around the epochs $t_2$ and $t_3$. Notations are same as 
those in Fig.~\ref{fig:one}.
}
\label{fig:two}
\end{figure}
% --------------------------------------------------------------

% Figure 3 ------------------------------------------------------
\begin{figure*}[t]
\centering
\includegraphics[width=11.5cm]{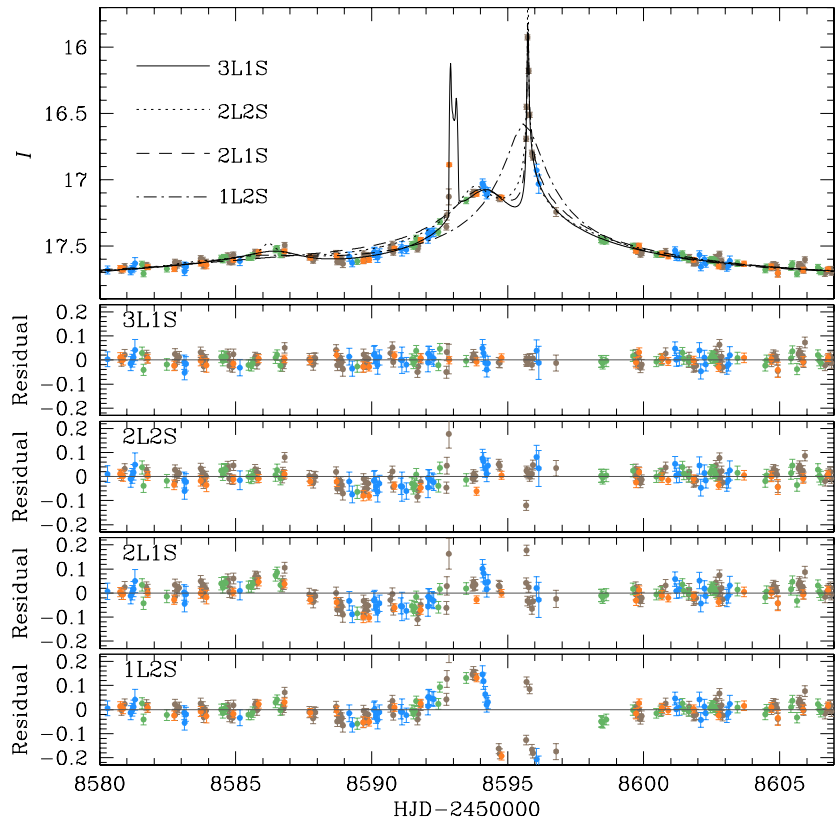}
\caption{
Comparison of four tested models, including 1L2S, 2L1S, 2S2L, and 3L1S (close-close solution) models.
The lower panels show the residuals from the individual models. 
}
\label{fig:three}
\end{figure*}
% --------------------------------------------------------------

Another group of target events of the project are those with known anomalies, but 
the interpretations of the anomalies have not been presented. From the investigation of such events, 
it was found that KMT-2019-BLG-1715 was a planetary event involved with three lens masses and two 
source stars \citep{Han2021c}, KMT-2019-BLG-0797 was a binary-lensing event occurring on a binary 
stellar system,  2L2S event \citep{Han2021b}, and KMT-2019-BLG-1953 was a strong candidate planetary 
event with a lens composed of two planets and the host \citep{Han2020a}.  The events in the latter 
group have a common characteristics that interpreting the lensing light curves of the events requires 
one to add extra source or lens components in modeling.

In this work, we present the result found from the reanalysis of the lensing event OGLE-2019-BLG-0468. 
The light curve of the event was previously investigated with 2L1S and 1L2S interpretations, but no 
plausible solution was suggested. From the reanalysis of the event based on more sophisticated models, 
we find that the event was produced by a triple-lens (3L1S) system, in which the lens is composed of 
two giant planets and their host star.

We present the analysis of the event according to the following organization.  In Sect.~\ref{sec:two}, 
we describe observations of the lensing event and the characteristics of the observed light 
curve.  In Sect.~\ref{sec:three}, we depict various models tested to explain the observed light curve, 
including 2L1S, 1L2S, 2L2S, and 3L1S models.  In Sect.~\ref{sec:four}, we characterize the source star 
and estimate the angular Einstein radius.  In Sect.~\ref{sec:five}, we estimate the physical lens 
parameters using the available observables of the event.  In Sect.~\ref{sec:six}, we discuss the 
possibility that the baseline object is the lens.  We then summarize results and conclude in 
Sect.~\ref{sec:seven}.

\section{Observations and data}\label{sec:two}

The source of the lensing event OGLE-2019-BLG-0468 lies in the Galactic bulge field at the equatorial
coordinates $({\rm RA}, {\rm DEC})_{\rm J2000}=($17:45:37.44, $-24$:26:50.2), which correspond to the 
galactic coordinates $(l, b)=(3^\circ\hskip-2pt .834,\ 2^\circ\hskip-2pt .336)$. The apparent baseline 
$I$-band magnitude of the source is $I_{\rm base}=17.8$ according to the OGLE-IV photometry system.  As 
we will show in Sect.~\ref{sec:four}, the source is much fainter than the baseline magnitude and the 
baseline flux comes mostly from a blend.

Figure~\ref{fig:one} shows the light curve of OGLE-2019-BLG-0468. The rising of the source flux induced 
by lensing was first found by the Optical Gravitational Lensing Experiment IV \citep[OGLE-IV:][]{Udalski2015} 
survey in the early part of the 2019 season on 2019-04-13 (${\rm HJD}^\prime\equiv {\rm HJD}-2450000\sim 8586$).  
The OGLE team utilizes the 1.3~m telescope at the Las Campanas Observatory in Chile, and it is equipped with 
a camera yielding 1.4~deg$^2$ field of view.  The source flux increased until $t_1\sim 8586.3$, and then declined 
during $8586\lesssim {\rm HJD}^\prime \lesssim 8590$, producing a weak bump at around $t_1$.  The flux suddenly 
increased at $t_2\sim 8592.9$, suggesting that the source crossed a caustic induced by the multiplicity of the 
lens.  The detailed structure of the light curve for the three nights during $8596\lesssim {\rm HJD}^\prime 
\lesssim 8598$ could not be delineated because the OGLE observation of the event was not conducted during this 
period.  When the event was observed by the OGLE survey again on ${\rm HJD}^\prime \sim 8599$, the source 
flux continued to decline until it reached the baseline.

% Table 1 ------------------------------------------------
\begin{table*}[t]
\small
%\centering
\caption{Lensing parameters of 1L2S, 2L1S and 2L2S models\label{table:one}}
%\begin{tabular}{\columnwidth}{@{\extracolsep{\fill}}lll}
\begin{tabular}{lllll}
\hline\hline
\multicolumn{1}{c}{Parameter}     &
\multicolumn{1}{c}{2L1S}          &
\multicolumn{1}{c}{1L2S}          &
\multicolumn{1}{c}{2L2S}          \\
\hline
$\chi^2$                    &  2244.3                    &   4018.8                    &  1949.1                 \\    
$t_{0,1}$ (HJD$^\prime$)    &  $8594.721 \pm 0.024$      &   $8595.614 \pm 0.013$      &  $8585.635 \pm 0.124$   \\
$u_{0,1}$                   &  $0.060 \pm 0.003$         &   $0.0020 \pm 0.0001$       &  $0.0084 \pm 0.0014$    \\
$t_{0,2}$ (HJD$^\prime$)    &  --                        &   $8585.745 \pm 0.149$      &  $8595.361 \pm 0.022$   \\
$u_{0,2}$                   &  --                        &   $0.0070 \pm 0.0009$       &  $0.0053 \pm 0.0003$    \\
$\te$ (days)                &  $29.10 \pm 1.68$          &   $297.36 \pm 1.69$         &  $181.05 \pm 9.90$      \\
$s$                         &  $0.623 \pm 0.008$         &   --                        &  $0.110 \pm 0.004$      \\
$q$                         &  $0.067 \pm 0.002$         &   --                        &  $1.252 \pm 0.307$      \\
$\alpha$ (rad)              &  $2.257 \pm 0.010$         &   --                        &  $1.191 \pm 0.014$      \\
$\rho_1$ ($10^{-3}$)        &  --                        &   --                        &  --                     \\ 
$\rho_2$ ($10^{-3}$)        &  --                        &   --                        &  --                     \\
$q_F$                       &  --                        &   $0.22 \pm 0.02$           &  $7.15 \pm 0.44$        \\
\hline
\end{tabular}
\tablefoot{ ${\rm HJD}^\prime\equiv {\rm HJD}-2450000$.  
}
\end{table*}
% --------------------------------------------------------

The event was also located in the field covered by the Korea Microlensing Telescope Network survey 
\citep[KMTNet:][]{Kim2016}.  The KMTNet group utilizes three identical telescopes, each of which has 
a 1.6~m aperture and is mounted with a camera yielding 4 deg$^2$ field of view.  For the continuous 
coverage of lensing events, the telescopes are distributed in the three continents of the Southern 
Hemisphere, at the Siding Spring Observatory in Australia (KMTA), the Cerro Tololo Interamerican 
Observatory in Chile (KMTC), and the South African Astronomical Observatory in South Africa (KMTS).  
For both OGLE and KMTNet surveys, observations of the event were done mainly in the $I$ band, and 
a fraction of $V$-band images were acquired for the measurement of the source color.  The event was 
identified by the KMTNet survey from the post-season inspection of the 2019 season data, and  it was 
labeled as KMT-2019-BLG-2696. Hereafter we use only the OGLE event number, according to the order of 
discovery, to designate the event.  The light curve constructed with the additional KMTNet data revealed 
that there existed an additional peak at $t_3\sim 8595.7$, that was not covered by the OGLE data.  See 
Figure~\ref{fig:two} showing the zoom-in view of the light curve around the epochs $t_2$ and $t_3$.

The light curve of the event was constructed by conducting photometry of the source using the pipelines 
of the individual survey groups: \citet{Wozniak2000} for OGLE and \citet{Albrow2009} for KMTNet.  
In order to consider the scatter of the data and to make $\chi^2$ per degree of freedom for each data 
set unity, the error bars estimated by the pipelines were readjusted according to the prescription 
depicted in \citet{Yee2012}.

In addition to the photometric data, we also obtained two spectra, with a 1000 second exposure for 
each, of the baseline object on the night of 2021-06-03 (${\rm HJD}^\prime = 9398$), which is 
$\sim 2.2$~yrs after the event, using the Robert Stobie Spectrograph \citep{Burgh2003} mounted on 
the South African Large Telescope \citep[SALT,][]{Buckley2006}.  The spectroscopic data were reduced 
using a custom pipeline based on the PySALT package \citep{Crawford2010}, which accounts for basic 
CCD characteristics, removal of cosmic rays, wavelength calibration, and relative flux calibration.  
To estimate the stellar parameters ($T_{\rm eff}$, $\log g$, and [Fe/H]), we interpolated the 
observed spectra in an empirical grid \citep{Du2019}, which was constructed with massive spectra 
of LAMOST.  Unfortunately, it was difficult to securely estimate the stellar parameters due to the 
low signal-to-noise ratios of the spectra.

\section{Light curve interpretation}\label{sec:three}

\subsection{2L1S, 1L2S, and 2L2S models}\label{sec:three-one}

We first model the observed light curve under the assumption of the lens (2L1S model) or source
(1L2S model) binarity, which is the most common cause of lensing light curve anomalies. The
lensing light curve of a single-lens single-source (1L1S) event is characterized by three parameters
of $(t_0, u_0, \te)$, which represent the time of the closest lens-source approach, the lens-source
separation at $t_0$, and the event timescale, respectively. Adding an extra lens or source component
in modeling requires one to include extra parameters. For a 2L1S event, these extra parameters are 
$(s, q, \alpha)$, which denote the binary separation, mass ratio between the lens components, and 
the angle between the direction of the source motion and the binary axis (source trajectory angle), 
respectively. For a 1L2S event, the extra parameters are $(t_{0,2}, u_{0,2}, q_F)$, in which the first 
two are the closest approach time and separation of the source companion, and the last parameter 
indicates the flux ratio between the companion ($S_2$) and primary ($S_1$) source stars.  In all cases 
of the tested models, we include an additional parameter $\rho$, which denotes the ratio of the angular 
source radius $\theta_*$ to the angular Einstein radius $\thetae$, that is, $\rho=\theta_*/\thetae$ 
(normalized source radius), to account for possible finite-source effects in the lensing light curve. 
To distinguish parameters related to $S_1$ and $S_2$, we use the notations $(t_{0,1}, u_{0,1}, \rho_1)$ 
and $(t_{0,2}, u_{0,2}, \rho_2)$, respectively.

% Figure 4 ------------------------------------------------------
\begin{figure}[t]
\includegraphics[width=\columnwidth]{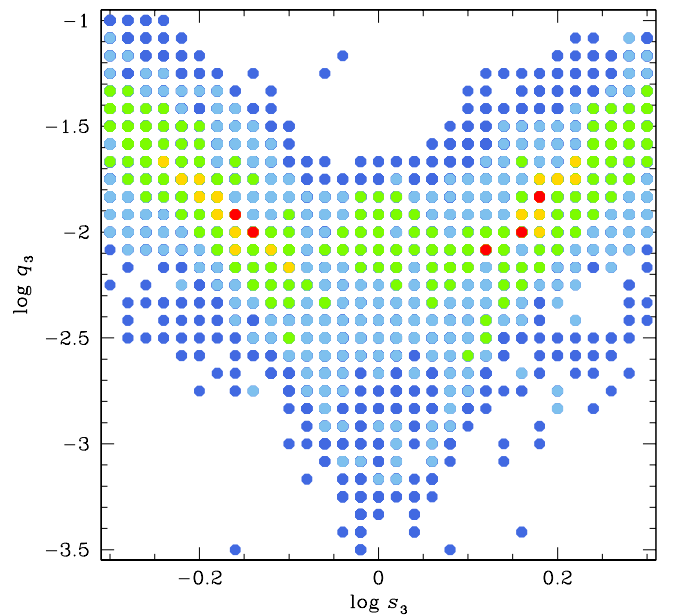}
\caption{
Distribution of $\Delta\chi^2$ in the $\log s_3$--$\log q_3$ parameter plane obtained from the 
preliminary 3L1S grid searches for the parameters related to the third lens component.  Red, yellow, 
green, cyan, and blues colors are used to denote points with $\Delta\chi^2 < n(1^2)$, $< n(2^2)$, 
$< n(3^2)$, $< n(4^2)$, and $< n(5^2)$, respectively, where $n=10$.
}
\label{fig:four}
\end{figure}
% --------------------------------------------------------------

From the modeling of the observed light curve with the 2L1S and 1L2S interpretations, it was found
that the data cannot be explained by these models. In Figure~\ref{fig:three}, we plot the model 
curves and residuals of the 2L1S (dashed curve) and 1L2S (dot-dashed curve) models. The lensing 
parameters of these models are listed in Table~\ref{table:one} together with the $\chi^2$ values 
of the fits.

We additionally check a 2L2S model, in which both the lens and source are binaries, for example, 
MOA-2010-BLG-117 \citep{Bennett2018}, OGLE-2016-BLG-1003 \citep{Jung2017},  KMT-2019-BLG-0797 
\citep{Han2021b}, and  KMT-2018-BLG-1743 \citep{Han2021a},   The model curve and residual from 
the 2L2S solution are presented in Figure~\ref{fig:three}, and the lensing parameters of the 
solution are listed in Table~\ref{table:one}.  This model provides a better fit than the 2L1S 
and 1L2S models by $\Delta\chi^2 =295.2$ and 2069.7, respectively.  However, the model still 
leaves substantial residuals, indicating that a new interpretation of the light curve is needed.

% Figure 5 ------------------------------------------------------
\begin{figure}[t]
\includegraphics[width=\columnwidth]{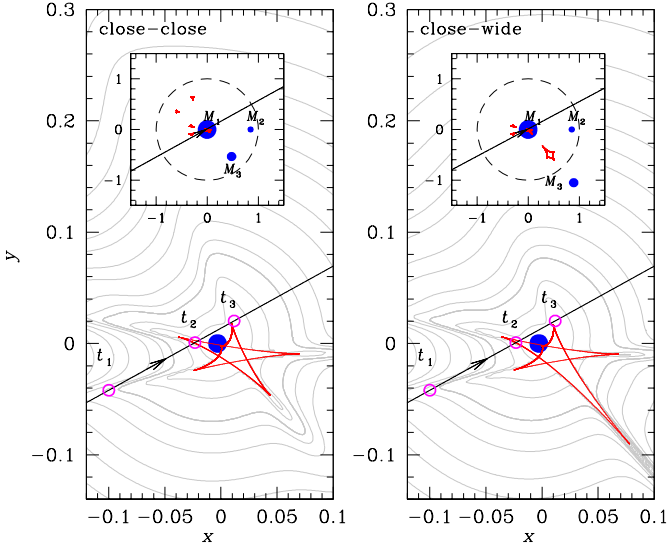}
\caption{
Lens system configurations of the close-close (left panel) and close-wide (right panel) 3L1L models.  
The inset in each panel is inserted to show the locations of the three lens components, denoted by 
blue dots marked by $M_1$, $M_2$, and $M_3$.  The dashed circle in the inset represents the Einstein 
ring. The line with an arrow represents the source trajectory. The three small empty dots on the source 
trajectory represent the source positions at the epochs of $t_1$, $t_2$, and $t_3$, that are marked in 
Fig.~\ref{fig:one}. The size of the dots is not scaled to the source size.  The red figure represents 
the caustic, and the grey curves around the caustic represent equi-magnification contours.  
}
\label{fig:five}
\end{figure}
% --------------------------------------------------------------

\subsection{3L1S model}\label{sec:three-two}

We then test a model in which the lens is a triple system (3L1S model).  In the first step of this 
modeling, we check whether a 2L1S model can describe part of the anomalies, although it turned 
out that the model could not explain all anomaly features.  We do this check because lensing light 
curves with three lens components ($M_1$, $M_2$, and $M_3$) can, in many cases, be approximated by 
the superposition of two 2L1S light curves produced by $M_1$--$M_2$ and $M_1$--$M_3$ lens pairs 
\citep{Bozza1999, Han2001}.  See example light curves of OGLE-2012-BLG-0026 \citep{Han2013}, generated 
by a lens composed of two planets and a host star, and OGLE-2018-BLG-1700 \citep{Han2020c}, produced 
by a lens composed of a planet and binary stars. From the modeling conducted with the partial data 
excluding those after $t_2$, it is found that the light curve is well approximated by a 2L1S model 
with binary parameters $(s, q, \alpha)\sim (0.86, 3\times 10^{-3}, 154^\circ)$. This suggests the 
possibility that the event may be produced by a 3L system.

% Figure 6 ------------------------------------------------------
\begin{figure}[t]
\includegraphics[width=\columnwidth]{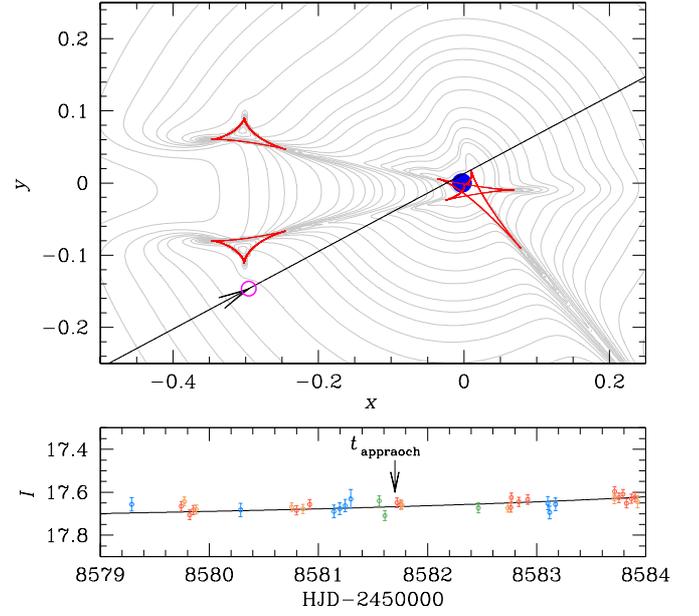}
\caption{
The magnification pattern around the caustics (for the close-wide 3L1S model) and the light curve 
around the time of the source approach close to the planetary caustic induced by $M_2$.  The small 
empty circle on the source trajectory represents the source position at the time of the source 
approach, at $t_{\rm approach} \sim 8581.7$ in ${\rm HJD}^\prime$.  The size of the circle is 
arbitrarily set, and is not scaled to the source size.
}
\label{fig:six}
\end{figure}
% --------------------------------------------------------------

% Figure 7 ------------------------------------------------------
\begin{figure}[t]
\includegraphics[width=\columnwidth]{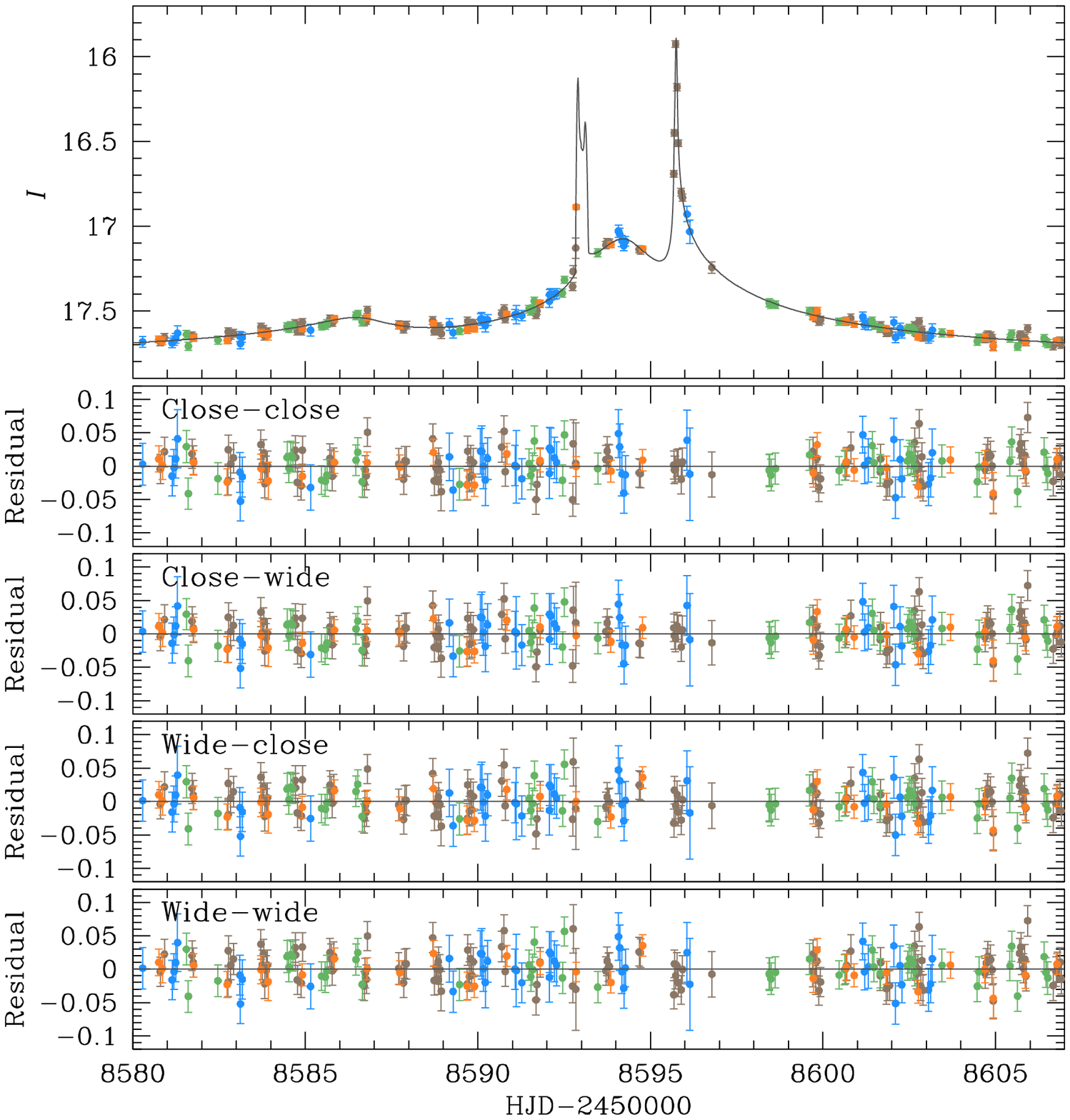}
\caption{
Comparison of the residuals from the four degenerate 3L1S solutions: close-close, close-wide, 
wide-close, and wide-wide solutions.  As a representative model, the model curve of the close-close 
solution is drawn over the data points  in the top panel. 
}
\label{fig:seven}
\end{figure}
% --------------------------------------------------------------

Modeling with the inclusion of a third lens component requires one to add three extra parameters to 
those of the 2L1S model. These parameters are $(s_3, q_3, \psi)$, which represent the $M_1$--$M_3$ 
separation, $M_3/M_1$ mass ratio, and the orientation angle of $M_3$ as measured from the $M_1$--$M_2$ 
axis with the center at the position of $M_1$, respectively.  In the first round of the 3L1S modeling, 
we search for the lensing parameters related to $M_3$, that is, $s_3$, $q_3$, and $\psi$, with a grid 
approach by fixing the parameters related to $M_2$, that is, $s_2$, $q_2$, and $\alpha$, as the ones 
obtained from the preliminary 2L1S modeling conducted with the use of the partial data.  
Figure~\ref{fig:four} shows the $\Delta\chi^2$ distribution in the $\log s_3$--$\log q_3$ parameter 
plane obtained from this first-round modeling.  We note that there exist two locals in the $\Delta\chi^2$ 
map, indicating that there are degenerate solutions: at $(\log s_3, \log q_3)\sim (+0.15, -2.0)$ and 
$\sim (-0.15, -2.0)$.  We will mention more details about the degeneracy below. In the second round, 
we refine the local solutions found from the first-round modeling using a downhill approach based on 
the Markov Chain Monte Carlo (MCMC) method. In this process, we release all parameters as free parameters.

We find that the observed  light curve including all three anomaly features is well explained by a 3L1S 
model.  The model curve is plotted over the data points in Figures~\ref{fig:one} and \ref{fig:two}, and 
the residual from the model is compared to those of the other models (1L2S, 2L1S, and 2L2S models) in 
Figure~\ref{fig:three}.  We find two solutions, in which $s_2<1.0$ and $s_3<1.0$ for one solution, and 
$s_2<1.0$ and $s_3>1.0$ for the other solution.  We refer to the individual solutions as ``close-close'' 
and ``close-wide'' solutions, respectively.

% Table 2 ------------------------------------------------
\begin{table*}[t]
\small
%\centering
\caption{Lensing parameters of four degenerate 3L1S models\label{table:two}}
%\begin{tabular}{\columnwidth}{@{\extracolsep{\fill}}lll}
\begin{tabular}{lllll}
\hline\hline
\multicolumn{1}{c}{Parameter}      &
\multicolumn{1}{c}{close-close}    &
\multicolumn{1}{c}{close-wide}     &
\multicolumn{1}{c}{wide-close}     &
\multicolumn{1}{c}{wide-wide}      \\
\hline
$\chi^2$                 &   1268.8                  &  1269.4                  &   1282.7                  &  1284.3                \\    
$t_0$ (HJD$^\prime$)     &   $8594.422 \pm 0.020$    &  $8594.411 \pm 0.020$    &   $8594.304 \pm 0.024$    &  $8594.301 \pm 0.023$  \\
$u_0$ ($10^{-3}$)        &   $11.98 \pm 0.45    $    &  $11.34 \pm 0.53    $    &   $12.73 \pm 0.54    $    &  $12.38 \pm 0.77    $  \\
$\te$ (days)             &   $75.24 \pm 2.02    $    &  $76.92 \pm 2.89    $    &   $76.24 \pm 2.61    $    &  $77.92 \pm 3.56    $  \\
$s_2$                    &   $0.853 \pm 0.003   $    &  $0.854 \pm 0.004   $    &   $1.107 \pm 0.005   $    &  $1.090 \pm 0.006   $  \\
$q_2$ ($10^{-3}$)        &   $3.54 \pm 0.18     $    &  $3.31 \pm 0.20     $    &   $4.26 \pm 0.18     $    &  $3.92 \pm 0.28     $  \\
$\alpha$  (rad)          &   $2.633 \pm 0.010   $    &  $2.648 \pm 0.010   $    &   $2.563 \pm 0.009   $    &  $2.567 \pm 0.009   $  \\
$s_3$                    &   $0.717 \pm 0.007   $    &  $1.379 \pm 0.012   $    &   $0.680 \pm 0.006   $    &  $1.479 \pm 0.016   $  \\
$q_3$ ($10^{-3}$)        &   $10.56 \pm 0.48    $    &  $10.47 \pm 0.51    $    &   $13.39 \pm 0.61    $    &  $14.13 \pm 0.96    $  \\
$\psi$  (rad)            &   $5.445 \pm 0.028   $    &  $5.416 \pm 0.023   $    &   $5.590 \pm 0.019   $    &  $5.622 \pm 0.019   $  \\ 
$\rho$ ($10^{-3}$)       &   $0.52 \pm 0.03     $    &  $0.50 \pm 0.03     $    &   $0.43 \pm 0.04     $    &  $0.43 \pm 0.05     $  \\
$f_s$                    &   $0.029 \pm 0.001   $    &  $0.028 \pm 0.001   $    &   $0.030 \pm 0.001   $    &  $0.029 \pm 0.002   $  \\
$f_b$                    &   $1.163 \pm 0.0     $    &  $1.164 \pm 0.001   $    &   $1.161 \pm 0.001   $    &  $1.162 \pm 0.001   $  \\ 
\hline
\end{tabular}
%\tablefoot{ ${\rm HJD}^\prime\equiv {\rm HJD}-2450000$.  }
\end{table*}
% --------------------------------------------------------

The lensing parameters of the two 3L1S solutions are listed in Table~\ref{table:two}. Also listed in 
the table are $\chi^2$ values of the fits, and the flux parameters of the source, $f_s$, and the blend, 
$f_b$, as measured on the OGLE flux scale.  We note that the estimated source flux, $f_s\sim 0.03$, 
is much smaller than the blend flux, $f_b\sim 1.16$, indicating that the observed flux is heavily 
blended.  The estimated mass ratios of the companions to the primary lens are $q_2\sim (3.3-3.5)
\times 10^{-3}$ and $q_3\sim 10.5\times 10^{-3}$, both of which correspond to the ratio between a 
giant planet and a star.  This indicates that the lens is a planetary system composed of two giant 
planets.  We note that the designation of $q_2$ and $q_3$ is not the order of a mass, and it turns 
out that $q_2 < q_3$.  We find that the fit of the 3L1S model is better than those the 1L2S, 2L1S, 
and 2L2S models by $\Delta\chi^2 =2750.0$, 974.5, and 680.3, respectively.  The degeneracy between 
the close-close and close-wide solutions is very severe, and the close-close solution is preferred over 
the close-wide solution by merely $\Delta\chi^2=0.6$.  The fact that the $M_1$--$M_3$ separations of 
the two degenerate solutions are in the relation of $s_{\rm 3,close-close}\times  s_{\rm 3,close-wide}
\sim 1$ indicates that the degeneracy is caused by the close-wide degeneracy in $s_3$ \citep{Griest1998, 
Dominik1999}.

% Figure 8 ------------------------------------------------------
\begin{figure}[t]
\includegraphics[width=\columnwidth]{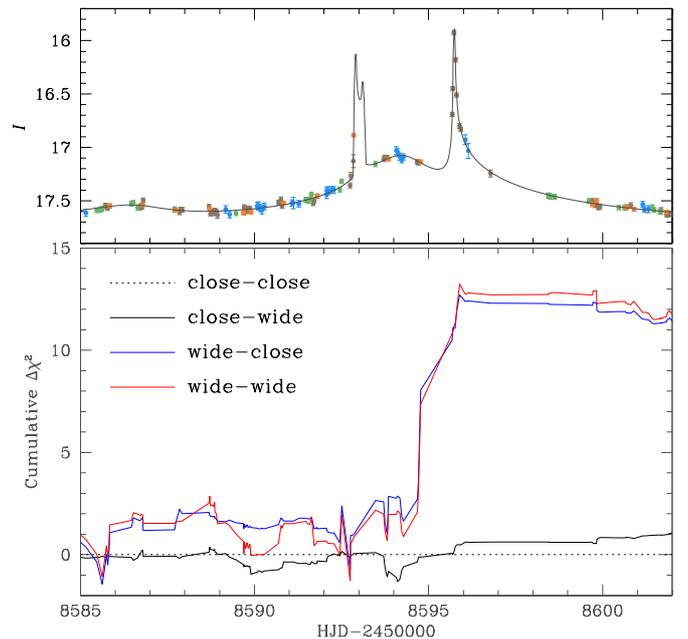}
\caption{
Cumulative distributions of $\Delta\chi^2$ difference from the best-fit solution (close-close model).
The light curve in the upper panel is inserted to show the locations of fit difference.
}
\label{fig:eight}
\end{figure}
% --------------------------------------------------------------

Figure~\ref{fig:five} shows the lens system configurations of the two 3L1S models.  For each model, 
the caustic (red figures) configuration appears to be the superposition of the two sets of caustics 
induced by two low-mass companions (blue dots marked by $M_2$ and $M_3$ in the inset of each panel) 
to the primary ($M_1$).  We mark the positions of the source at the three epochs of $t_1$, $t_2$, and 
$t_3$ (small empty circles drawn in magenta). The source position at $t_1$ corresponds to the 
source passage through the positive deviation region extending from one of the sharp cusp of the 
caustic induced by $M_2$, and this produced a weak bump at around $t_1$. The source then 
crossed the tip of the caustic, producing a sharp caustic-crossing feature, and the two data points 
at $t_2$, one from OGLE and the other from KMTC observations, correspond to the time of the caustic 
entrance.  Another caustic-crossing occurred when the source passed the tip of the caustic induced by 
$M_3$, and this produced a sharp peak around $t_3$. The last peak was covered by the KMTC data, both 
in the rising and falling sides of the peak.

The insets of the panels in Figure~\ref{fig:five} shows that the source passes close to one of 
the triangular planetary caustics induced by $M_2$. If the separation between the source and the 
caustic were very close, this source approach would induce a low bump at the time of the caustic 
approach. We check this possibility of the bump by inspecting the magnification pattern around the
planetary caustic.   The magnification map around the caustics (for the close-wide 3L1S model) is 
shown in Figure~\ref{fig:six}, in which the source location at the time of the caustic approach is 
marked by an empty magenta circle. The source size, $\rho\sim 0.5\times 10^{-3}$, is too small to 
be seen in the presented scale, and thus we arbitrarily set the circle size.  The lower panel shows 
the light curve around the time of the caustic approach at $t_{\rm approach}\sim 8581.7$.  Both the 
magnification map and the light curve show that the deviation at around this approach is too weak 
to emerge from the baseline.

The fact that the caustic configuration of the 3L1S solutions appears to be the superposition of two
2L1S caustics together with the fact that $M_3$ is in the planetary-mass regime suggest that there
may be additional degenerate solutions caused by the close-wide degeneracy in $s_2$. We, therefore,
search for additional solutions resulting from this degeneracy: solutions with $s_2>1.0$, $s_3<1.0$
(``wide-close'' solution) and $s_2>1.0$, $s_3>1.0$ (``wide-wide'' solution).  These degenerate solutions 
are found using the initial parameters of $s_{2,{\rm wide,xx}}=1/s_{2,{\rm close-xx}}$.  The lensing 
parameters of these solutions are listed in Table~\ref{table:two}. We find that, although the wide-xx 
solutions approximately describe the observed light curve, their fits are worse than the close-xx 
solutions by $\Delta\chi^2\sim 14$, that are mostly contributed by the 3 data points (one from OGLE and 
two from KMTC) taken at ${\rm HJD}^\prime \sim 8594.7$ and the 7 data points (from KMTC) covering the 
anomaly at $t_3$. The superiority of the close-xx solutions over the wide-xx solutions is found from the 
comparison of the residuals of the models, presented in Figure~\ref{fig:seven}, as well as the cumulative 
distributions of $\chi^2$ difference from the best-fit solution (close-close solution), presented in 
Figure~\ref{fig:eight}. The degeneracy in $s_2$ is resolvable because the separation ($s_2\sim 0.85$ 
for the close solution and $\sim 1.1$ for the wide solution) is close to unity. In this case, the 
$M_1$--$M_2$ pair forms a resonant caustic, for which the close-wide degeneracy is less severe 
\citep{Chung2005}.

% Figure 9 ------------------------------------------------------
\begin{figure}[t]
\includegraphics[width=\columnwidth]{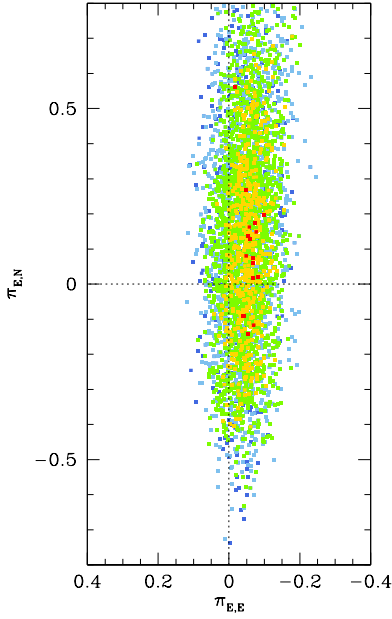}
\caption{
Scatter plot of points in the MCMC chain on the $\piee$--$\pien$ parameter plane for the close-close 
3L1S solution.  The color coding is same as that in Fig.~\ref{fig:four}, except that $n=1$.
}
\label{fig:nine}
\end{figure}
% --------------------------------------------------------------

We check whether the parameters of the normalized source radius, $\rho$, and the microlens-parallax, 
$\pie$, can be measured.  Measurements of these parameters are important because the mass, $M$, and 
distance to the lens, $\dl$, can be uniquely determined with these parameters by 
\begin{equation}
M= {\thetae \over \kappa\pie };\qquad
\dl = {{\rm AU} \over \pie\thetae+ \pi_{\rm S} };\qquad
\thetae = {\theta_*\over \rho},
\label{eq1}
\end{equation}
where $\kappa= 4G/(c^2{\rm AU})$ and $\pi_{\rm S}$ denotes the annual parallax of the source
\citep{Gould2000}.  The value of the normalized source radius, $\rho \sim 0.5\times 10^{-3}$, is 
firmly measured from the data points involved with caustic crossings at around $t_2$ and $t_3$.  On 
the other hand, modeling considering the microlens-parallax effect results in a $\pie$ value with a 
considerable uncertainty due to fact that the photometry quality of the data is mediocre because of 
the faintness of the source, together with that all main features occur in an interval of 12 days, 
which is too short to appreciate any deviations due to parallax effects.  Figure~\ref{fig:nine} 
shows the scatter plot of points in the MCMC chain on the $\piee$--$\pien$ plane, where $\piee$ 
and $\pien$ are the east and north components of the microlens-parallax vector $\pivec_{\rm E}$, 
respectively.  The scatter plot shows that the microlens-parallax is consistent with a zero-parallax 
model within $2\sigma$ level, and the uncertainty of $\pien$ is big, although the uncertainty of 
$\piee$ is relatively small.  Considering that the anomaly features in the light curve occurred 
within a short time interval, the lens-orbital motion would not invoke any significant effects.  
Nevertheless, we conduct an additional modeling considering the lens-orbital model, and found 
that the fit improvement by the lens-orbital effect is negligible and the orbital motion was 
poorly constrained.

\section{Source star and angular Einstein radius}\label{sec:four}

For the determination of the angular Einstein radius, it is required to estimate the angular 
radius of the source. We deduce $\theta_*$ from the color and brightness of the source.  To 
estimate the extinction- and reddening-corrected (dereddened) values, $(V-I, I)_0$, from the 
instrumental values, $(V-I, I)$, we use the \citet{Yoo2004} method, in which the centroid of 
red giant clump (RGC), $(V-I, I)_{\rm RGC}$, in the color-magnitude diagram (CMD) is used for 
the calibration of the source color and magnitude.

\begin{figure}[t]
\includegraphics[width=\columnwidth]{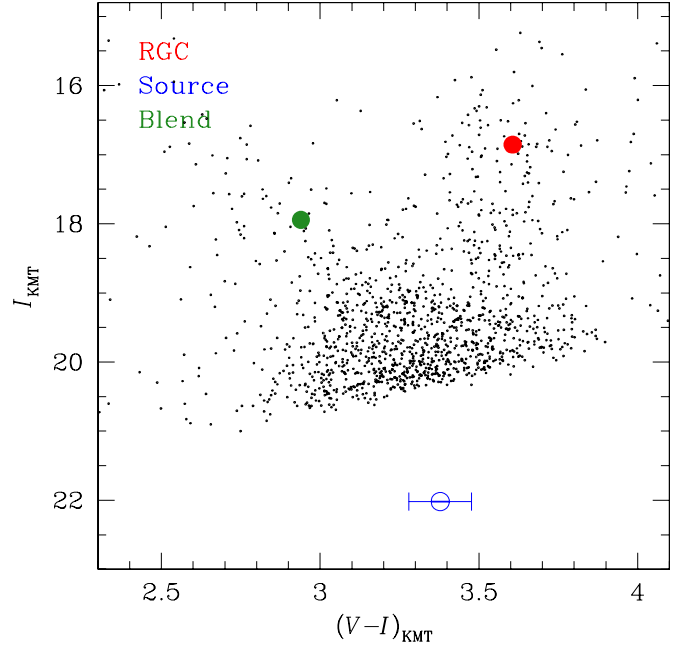}
\caption{
Locations of the source (blue empty dot with error bars), RGC centroid (red filled dot), and 
blend (green filled dot) in the instrumental color-magnitude diagram of stars around the source 
constructed using the pyDIA photometry of the KMTC data set.  
}
\label{fig:ten}
\end{figure}

Figure~\ref{fig:ten} shows the locations of the source (blue empty dot with error bars) and the 
RGC centroid (red filled dot) in the instrumental CMD constructed using the KMTC data processed  
using the pyDIA software \citep{Albrow2017}.  The source position in the CMD is determined by the 
regression of the $I$- and $V$-band pyDIA data with the variation of the lensing magnification.  
Also marked in the CMD is the blend position (green filled dot), which lies on the main-sequence 
branch of foreground disk stars.  We show in Sect.~\ref{sec:six} that the blended light is due to
either the host star, a companion to the host star, or a combination of the two.  The measured 
instrumental color and brightness are $(V-I, I) = (3.378\pm 0.099, 22.018\pm 0.004)$ for the source 
and $(V-I, I)_{\rm RGC} = (3.606, 16.856)$ for the RGC centroid. By measuring the offsets in color, 
$\Delta (V-I)$, and magnitude, $\Delta I$, between the source and RGC centroid, together with the 
known dereddened values of the RGC centroid, $(V-I,I)_{\rm RGC,0}=(1.060, 14.268)$, from 
\citet{Bensby2013} and \citet{Nataf2013}, respectively, we estimate that the dereddened source 
color and magnitude are
\begin{equation}
\eqalign{
(V-I, I)_0 & =  (V-I, I)_{{\rm RGC},0 }+ \Delta (V-I, I) \cr
           & =  (0.832\pm 0.099, 19.431\pm 0.004).       \cr
}
\label{eq2}
\end{equation}
This indicates that the source is a late G-type main-sequence star located in the bulge.

% Figure 10 ------------------------------------------------------
% --------------------------------------------------------------

For the $\theta_*$ estimation from $(V-I, I)_0$, we first convert the measured $V-I$ color into 
$V-K$ color using the color-color relation of \citet{Bessell1988}, and second deduce the source 
radius from the $(V-K)$--$\theta_*$ relation of \citet{Kervella2004}. The estimated value is
\begin{equation}
\theta_* = 0.47 \pm 0.06~\mu{\rm as}.
\label{eq3}
\end{equation}
Together with the $\rho$ and $\te$ values measured from modeling, the angular Einstein radius 
and the relative lens-source proper motion are estimated as
\begin{equation}
\thetae = 0.91  \pm 0.13~{\rm mas},
\label{eq4}
\end{equation}
and
\begin{equation}
\mu = 4.41 \pm 0.63~{\rm mas}~{\rm yr}^{-1}, 
\label{eq5}
\end{equation}
respectively.

% Figure 11 ------------------------------------------------------
\begin{figure}[t]
\includegraphics[width=\columnwidth]{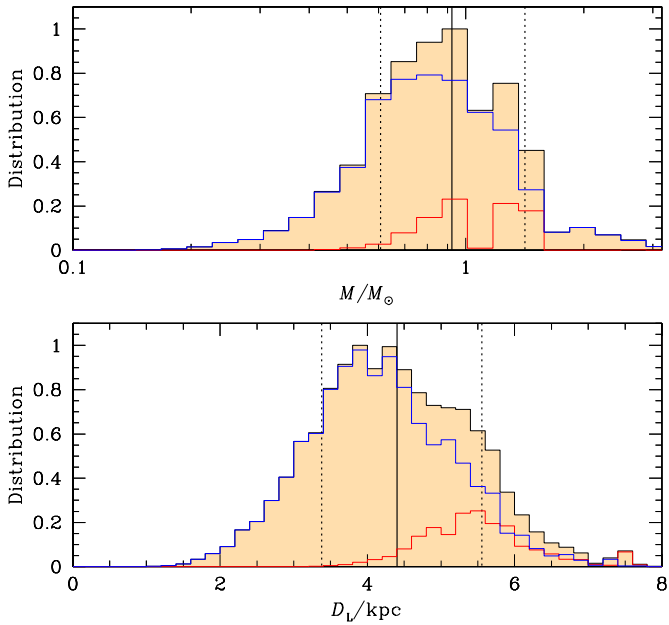}
\caption{
Bayesian posterior distributions of the mass (upper panel) and distance (lower panel) to the lens. 
In each panel, the solid vertical line indicates the median of the distribution, and the two dotted 
vertical lines indicate the 1$\sigma$ range of the distribution. The blue and red curves represent 
the contributions by disk and bulge lenses, respectively.  
}
\label{fig:eleven}
\end{figure}
% --------------------------------------------------------------

\section{Physical lens parameters}\label{sec:five}

Although the mass and distance to the lens cannot be uniquely determined using the relation in
Equation~(\ref{eq1}) due to the insecure measurement of the microlens parallax, the physical 
parameters can still be constrained using the measured observables of $\te$ and $\thetae$, which 
are related to $M$ and $\dl$ by
\begin{equation}
\te={\thetae \over \mu};\qquad
\thetae = (\kappa M \pi_{\rm rel})^{1/2}.
\label{eq6}
\end{equation}
Here $\pi_{\rm rel}={\rm AU}(D_{\rm L}^{-1}-D_{\rm S}^{-1})$ represents the relative lens-source 
parallax and $\ds$ denotes the distance to the source.  For this constraint, we conduct a Bayesian 
analysis using a prior Galactic model.  Although the uncertainty of the measured $\pie$ is big, we 
consider the measured $\pie$ by computing the parallax-ellipse covariance matrix in the Bayesian 
estimation of the physical lens parameters.

The Galactic model defines the physical and dynamical distributions and the mass function of Galactic 
objects.  We adopt the Galactic model described in \citet{Jung2021}. The model uses the \citet{Robin2003} 
and \citet{Han2003} physical matter distributions to specify the locations of disk and bulge objects, 
respectively. It also uses the \citet{Han1995} and \citet{Jung2021} dynamical distributions to describe 
the motions of disk and bulge objects, respectively. The mass function adopted in the model is described 
in \citet{Jung2018}, and it includes stellar remnants and brown dwarfs.

In the first step of the Bayesian analysis, we conduct a Monte Carlo simulation to produce many ($10^7$)
artificial lensing events, for which lens and source locations, transverse lens-source speeds, and lens 
masses are allocated following the Galactic model. Then, the microlensing observables $(\te, \thetae, \pie)$ 
of the individual events are computed using the relations in Equations~(\ref{eq1}) and (\ref{eq6}). In the 
second step, we construct the distributions of $M$ and $\dl$ of events for which their observables are 
consistent with the observed values.

Figure~\ref{fig:eleven} shows the posterior distributions of $M$ and $\dl$ obtained from the Bayesian analysis. In
Table~\ref{table:three}, we list the estimated physical parameters of the masses of the individual lens components
$(M_1, M_2, M_3)$, distance, and the projected physical separations $(a_{\perp,2}, a_{\perp,3})$ between $M_1$--$M_2$ 
and $M_1$--$M_3$ pairs for the close-close and close-wide solutions.  We adopt the median values of the distributions 
as representative parameters, and the uncertainties of the parameters are estimated as 16\% and 84\% of the distributions.  
The two solutions result in similar physical parameters except for $a_{\perp,3}$. The estimated mass of the host,
\begin{equation}
M_1 = 0.92^{+0.49}_{-0.32}~M_\odot,
\label{eq7}
\end{equation}
indicates that the host is a G-type star, and the distance, 
\begin{equation}
\dl = 4.40^{+1.15}_{-1.02}~{\rm kpc},
\label{eq8}
\end{equation}
indicates that the lens is likely to be in the disk.  The two planets have masses
\begin{equation}
(M_2, M_3)=
\begin{cases}
(3.43^{+1.83}_{-1.17}~M_{\rm J}, 10.22^{+5.46}_{-3.50}~M_{\rm J}), & \text{for close-close},\\
(3.21^{+1.71}_{-1.10}~M_{\rm J}, 10.14^{+5.42}_{-3.47}~M_{\rm J}), & \text{for close-wide},
\end{cases}
\label{eq9}
\end{equation}
and thus both planets are heavier than Jupiter.  Considering that the snow line distance from the host is 
$ d_{\rm sl}= 2.7(M/M_\odot)\sim 2.5~{\rm AU}$, the estimated separations, 
\begin{equation}
(a_{\perp,2}, a_{\perp,3})=
\begin{cases}
(3.29^{+4.15}_{-2.52}~{\rm AU}, 2.77^{+3.49}_{-2.12}~{\rm AU}),  &  \text{for close-close}, \\
(3.29^{+4.16}_{-2.53}~{\rm AU}, 5.32^{+6.71}_{-4.08}~{\rm AU}),  &  \text{for close-wide},  
\end{cases}
\label{eq10}
\end{equation}
indicate that both planets are located at around and slightly beyond the snow line, similar to the giant 
planets in the solar system.

% Table 3 ------------------------------------------------
\begin{table}[t]
\small
%\centering
\caption{Physical lens parameters\label{table:three}}
%\begin{tabular}{llll}
\begin{tabular*}{\columnwidth}{@{\extracolsep{\fill}}lccc}
\hline\hline
\multicolumn{1}{c}{Parameter}         &
\multicolumn{1}{c}{close-close}       &
\multicolumn{1}{c}{close-wide}        \\
\hline
$M_1$ ($M_\odot$)      &   $0.92^{+0.49}_{-0.32} $  &  $\leftarrow$            \\
$M_2$ ($M_{\rm J}$)    &   $3.43^{+1.83}_{-1.17} $  &  $3.21^{+1.71}_{-1.10} $ \\
$M_3$ ($M_{\rm J}$)    &   $10.22^{+5.46}_{-3.50}$  &  $10.14^{+5.42}_{-3.47}$ \\
$\dl$ (kpc)            &   $4.40^{+1.15}_{-1.02} $  &  $\leftarrow$            \\
$a_{\perp,2}$ (AU)     &   $3.29^{+4.15}_{-2.52} $  &  $3.29^{+4.16}_{-2.53} $ \\
$a_{\perp,3}$ (AU)     &   $2.77^{+3.49}_{-2.12} $  &  $5.32^{+6.71}_{-4.08} $ \\
\hline
\end{tabular*}
%\end{tabular}
\tablefoot{ 
The arrows in the right column imply that the values are same as those in the middle column.
}
\end{table}

\section{Baseline object}\label{sec:six}

From the position of the blend on the CMD (Figure~\ref{fig:ten}), it is found that the flux of 
the baseline object is dominated by the light from a main-sequence star in the foreground disk.  
Logically, there are only four possibilities for this star: (1) the host of the planets, (2) a 
companion to the host, (3) a companion to the source, (4) an ambient star that is unrelated to the 
event (or possibly a combination of two or more of these).  The position of the blend on the CMD 
already rules out the possibility (3) because it is inconsistent with a bulge star.

Because $\thetae\sim 0.9$~mas, the lens must also be in the foreground disk (unless it is a dark 
remnant).  That is, if the lens were in the bulge, with $\pi_{\rm rel}\la 0.02$~mas, then its mass 
would be $M=\thetae^2/\kappa\pi_{\rm rel}\gtrsim 5\,M_\odot$, implying that it would be easily seen.  
This suggests that the blended light may be due primarily to the host of the planets.

To test this conjecture we first measure the astrometric offset between the ``baseline object'' and 
the source (which has the same position as the host at the time of the event).  The position of the 
source is measured from difference images formed by subtracting the reference image from a series of 
images taken at high magnifications.  This difference image essentially consists of an isolated star 
on a blank background.  Hence, its position can be accurately measured.  We find a scatter 
among 15 images of just (0.026, 0.037) pixels in the (east, north) directions, yielding a standard 
error of the mean of (0.007, 0.010) pixels, that is, (3, 4)~mas, given the 400 mas pixel size.  These 
errors are far below the other errors in the problem, which are discussed below.  We therefore ignore 
them in what follows.

By contrast, the problem of measuring the position of the baseline object is far more subtle.  Its 
position is returned by the DoPhot photometry package \citep{Schechter1993}, which simultaneously 
fits for the positions and fluxes from possibly overlapping stellar images.  Fortunately, the baseline 
object appears isolated on these images, so that the statistical errors of this measurement are not 
strongly impacted by the algorithm's procedure for separating stars.  We estimate this error to be 
23~mas in each direction from a comparison of two completely independent reductions, that is, by 
different people.  We find an offset between the source and the baseline object is 
\begin{equation}
%\Delta\btheta(E,N) =(36, 73)\pm (23, 23)~{\rm mas}.  
\Delta\btheta(E,N) =(36\pm 23, 73\pm 23)~{\rm mas}.  
\label{eq11}
\end{equation}
If we could ignore systematic effects and take the error distributions 
to be a 2-dimensional Gaussian, this would rule out the identification of the host with baseline light 
at $p=0.002$ probability.

We note that this close alignment rules out possibility (4), that is, that the blended light is 
due to an ambient star.  We simply count the number of stars on the foreground main sequence in the 
2~arcmin$^2$ area of the OGLE finding chart that are brighter than the baseline object, finding $N=64$.  
This translates to a surface density of $n=64/120^2 = 4.4 \times 10^{-3}\,{\rm arcsec}^{-2}$.  Then 
the probability of such an alignment is $p = \pi (\Delta\theta)^2 n = 10^{-4}$.  Hence, the light 
from the baseline object must be due to either the planet host or a companion to the host (or a 
combination).

Before considering these two remaining possibilities, we examine the role of the key systematic 
effect that could corrupt the position measurement of the baseline object at the $\Delta\theta=
[\Delta\theta(N)^2+ \Delta\theta(E)^2]^{1/2}\sim 80$~mas level: namely, the possibility that an 
ambient star lies within the point-spread function (PSF) of the baseline object.   That is, if a 
fraction $f$ of the blended light were due to an ambient star with a separation from the source of 
$\delta\btheta$, then the centroid of light (measured by DoPhot) would be displaced by $\Delta\btheta 
= f\delta\btheta$.  For instance, for $f=0.2$ and $\delta\theta=400$~mas, $\Delta\theta=80$~mas.  
In this example, the faint ambient star, with separation $0.4^{\prime\prime}$ would not be identified 
as a separate star by DoPhot, so the centroid would be shifted.

We have just argued that ambient stars similar to or brighter than the baseline object are rare.  
However, ambient stars that could corrupt the astrometric measurement at this level are not rare, 
for three reasons.  First, such stars are not restricted to the foreground main sequence, so the 
much larger population of bulge stars is available.  Second, there are more faint stars than bright 
stars.  Third, ambient stars can corrupt the astrometric measurement with separations up to 
$\delta\theta\sim 1^{\prime\prime}$, whereas ambient stars that would explain the baseline object 
light must lie within $\Delta\theta=80$~mas.  These three features also place limits on which ambient 
stars can play this role.  First, if the ambient star were too bright (and came from the more populous, 
but redder bulge population), then it would also drive the combined color of the resulting baseline 
object to the red, and so off the foreground main-sequence feature, contrary to its actual location.  
Second, if the ambient star were too faint, its separation would have to be $\delta\theta\ga 1^{\prime\prime}$, 
at which point its light would no longer enter into the DoPhot centroid.  Third, as $\delta\theta$ is 
increased, ambient stars of sufficient brightness will be separately resolved by DoPhot before the 
$\delta\theta=1^{\prime\prime}$ limit is reached.

To make our evaluation, we first restrict attention to the 2349 stars in the $2$ arcmin$^2$ OGLE-IV 
finding chart that satisfy $0.7<(I-I_{\rm base})<2.7$, where $I_{\rm base} = 17.8$ is evaluated in 
the OGLE-IV system.  The faint limit is set by the requirement $\delta\theta<1^{\prime\prime}$, while 
the bright limit is set by the requirement that inclusion of the (usually) red ambient light does not 
drive the baseline object off the foreground main sequence.  Then, for each of these 2349 stars we 
find the annulus of positions such that the star would induce as astrometric error $>80$~mas.  The 
inner radius of the annulus is set by the flux ratio: $\delta\theta_{\rm inner}=\Delta\theta\times
10^{0.4(I-I_{\rm base})}$.  We set $\delta\theta_{\rm outer}=0.2^{\prime\prime}[(I-I_{\rm base})+2.3]$.  
We then find a total area subtended by these 2349 annuli to be $2800\,{\rm arcsec}^2$, i.e., a fraction 
$p=19\%$ of the 2 arcmin$^2$.  That is, if the host were primarily responsible for the blended light, 
then there is a $p=19\%$ probability that an ambient star would corrupt the astrometric measurement 
by enough to account for the observed $\Delta\theta\sim 80$~mas offset.

On the other hand, it is also possible that the baseline object has a fainter companion that generated 
the main microlensing event and so serves as the host for the planets.  The main constraint on this 
scenario is that the blend, which is likely a G dwarf at several kpc, should have a widely separated 
companion at $\log(P/{\rm day})\sim 4$, with mass ratio $q_{\rm base}\ga 0.5$.  According to Table 7 
of \citet{Duquennoy1991}, about 25\% of G dwarfs have companions with $\log(P/{\rm day})> 4$ and mass 
ratio $q\geq 0.5$.

However, if the baseline object did have a companion in this parameter range, it is about equally 
likely that the baseline object served as the lens-host for the event while the companion generated 
the flux required to corrupt the astrometric measurement.

In brief, there are two channels for the baseline object to be the host, with the astrometric measurement 
corrupted either by an ambient star ($p=0.19$) or by a less luminous widely separated companion to the 
planet host ($p=0.25/2\sim 0.12$,) for a total of $p=0.31$.  By contrast, there is one channel for the 
host to be a companion of the baseline object, with probability $p=0.25/2 \sim 0.12$.  Therefore, the 
host of the planet is likely responsible for the light of the baseline object, but the host could also 
plausibly be a companion to the baseline object.  Considering the inconclusive nature of the blend 
together with the low quality of the spectra, we do not impose the constraint from the spectra on the 
estimation of the physical parameters presented in Table~\ref{table:three}.

\section{Summary and conclusion}\label{sec:seven}

We reported a multiple planetary system discovered from the analysis of the lensing event 
OGLE-2019-BLG-0468, that was reinvestigated in the project of reviewing the previous microlensing 
data collected in and before the 2019 season by the KMTNet survey.  The light curve, consisted of 
three distinctive anomaly features, of the event could not be explained by the usual 2L1S or 1L2S 
interpretation.  We found a solution explaining all anomaly features with a triple-lens interpretation, 
in which the lens is composed of two planets and their host, making the lens the fourth multi-planet 
system securely found by microlensing.  The lensing solution is subject to two-fold degeneracies 
caused by the ambiguity in estimating the separations of the planets from the host. One fold of 
the degeneracy is very severe, but the other was resolvable due to the resonant nature of the 
caustic induced by the second planet. The two planets have masses $\sim 3.4~M_{\rm J}$ and 
$\sim 10.2~M_{\rm J}$, and they are orbiting around a G-type star with a mass $\sim 0.9~M_\odot$ 
and a distance $\sim 4.4$~kpc.  It was found that the planet host was most likely responsible for 
the light of the baseline object, although the possibility for the host to be a companion to the 
baseline object could not be ruled out.

\begin{acknowledgements}
Work by C.H. was supported by the grants  of National Research Foundation of Korea 
(2020R1A4A2002885 and 2019R1A2C2085965).
% Subo 
S.D. acknowledges the science research grants from the China Manned Space Project with 
NO.~CMS-CSST-2021-A11.
% KMTNet
This research has made use of the KMTNet system operated by the Korea Astronomy and Space 
Science Institute (KASI) and the data were obtained at three host sites of CTIO in Chile, 
SAAO in South Africa, and SSO in Australia.
%% OGLE  
%The OGLE project has received funding from the National Science Centre, Poland, grant
%MAESTRO 2014/14/A/ST9/00121 to AU.
% SALT 
The observations using the SALT telescope were conducted under the transients follow up
programme 2018-2-LSP-001 (PI: D.B.), which is supported by Poland under grant no.~MNiSW
DIR/WK/2016/07. D.B. also acknowledges research support from the National Research Foundation.
M.G. is supported by the EU Horizon 2020 research and innovation programme under grant
agreement No.~101004719.
\end{acknowledgements}

\end{document}